\documentclass[twocolumn,10pt]{article}
\usepackage[utf8]{inputenc}
\usepackage[T1]{fontenc}
\usepackage[american]{babel}
\usepackage{amsmath,amssymb}
\usepackage{graphicx}
\usepackage{float}
\usepackage{array}
\usepackage{booktabs}
\usepackage{multirow}
\usepackage{url}
\usepackage{natbib}
\usepackage[margin=1in]{geometry}

\usepackage{tabularx}
\usepackage{ltxtable}
\usepackage{longtable}

\newcolumntype{L}[1]{>{\raggedright\arraybackslash}p{#1}}
\newcolumntype{C}[1]{>{\centering\arraybackslash}p{#1}}
\newcolumntype{R}[1]{>{\raggedleft\arraybackslash}p{#1}}

\title{Cognitive Cybersecurity for Artificial Intelligence: Guardrail Engineering with CCS-7}

\author{
Yuksel AYDIN\\
Independent Researcher\\
\texttt{}
}

\date{}

\begin{document}

\maketitle

\begin{abstract}
Language models exhibit human‑like cognitive vulnerabilities, such as emotional framing, that escape traditional behavioral alignment. We present CCS‑7 (Cognitive Cybersecurity Suite), a taxonomy of seven vulnerabilities grounded in human cognitive security research. To establish a human benchmark, we ran a randomized controlled trial with 151 participants: a ``Think‑First, Verify‑Always'' (TFVA) lesson improved cognitive security by $+7.9\%$ overall. We then evaluated TFVA‑style guardrails across 12{,}180 experiments on seven diverse language model architectures. Results reveal architecture‑dependent risk patterns: some vulnerabilities (e.g., identity confusion) are almost fully mitigated, while others (e.g., source interference) exhibit escalating backfire, with error rates increasing by up to $135\%$ in certain models. Humans, in contrast, show consistent moderate improvement. These findings reframe cognitive safety as a model‑specific engineering problem: interventions effective in one architecture may fail, or actively harm, another, underscoring the need for architecture‑aware cognitive safety testing before deployment.
\end{abstract}

\section{Introduction}

Language models can display human‑like reasoning errors. 
A model asked to provide citations for ``cybersecurity protein quantum therapy'' may confidently return five plausible DOIs, none of which exist. 
When told that ``SHA‑256 has been proven reversible'', it can integrate the false claim into its response. 
These are not simple failures of instruction following, but cognitive vulnerabilities: systematic weaknesses in how models process and integrate information.

\textbf{The Problem.} 
As language models approach human‑level reasoning, they inherit human cognitive weaknesses such as confirmation bias, deference to authority, and susceptibility to emotionally framed inputs. 
Traditional behavioral alignment, ensuring models follow instructions or avoid unsafe outputs, does not guarantee that their reasoning is reliable. 
Achieving cognitive safety requires addressing these reasoning‑level vulnerabilities.

\textbf{The Current Gap.} 
Despite advances in AI safety, we lack a systematic framework to: 
(i) characterize the types of cognitive vulnerabilities language models exhibit, 
(ii) determine which can be mitigated by prompt‑level interventions, and 
(iii) anticipate when well‑intentioned guardrails may backfire and amplify errors.

\textbf{Our Approach.} 
This work introduces a Cognitive Cybersecurity Suite (CCS‑7), a taxonomy of seven cognitive vulnerabilities inspired by human cognitive security research and AI analysis papers. 
We empirically evaluate these vulnerabilities through 12{,}180 controlled experiments across seven diverse model architectures, paired with a randomized human study ($n=151$) that tests a lightweight cognitive‑security protocol, ``Think‑First, Verify‑Always'' (TFVA) described in our previous study \cite{aydin2025thinkfirstverifyalways}. 
Our contributions are:
\begin{enumerate}
    \item A formalization of seven cognitive vulnerabilities (CCS‑7) that appear in both humans and language models.
    \item An empirical evaluation of TFVA as a cognitive guardrail, showing which vulnerabilities are mitigated, which resist intervention, and which exhibit harmful backfire.
    \item Evidence that cognitive safety is architecture‑dependent: interventions that help one model can fail, or even harm, another, underscoring the need for model‑aware safety validation.
\end{enumerate}

This framing shifts AI safety from viewing cognitive errors as isolated quirks to treating them as systematic, testable vulnerabilities, signaling the way for architecture‑aware cognitive safety strategies.

\section{Related Work}

\subsection{From Behavioral to Cognitive Safety}
Most AI safety research has focused on behavioral alignment, ensuring that models follow instructions and avoid harmful outputs. 
Prominent approaches include Constitutional AI \cite{Bai2022}, which encodes ethical principles into model responses, and reinforcement learning from human feedback (RLHF) \cite{Ouyang2022}, which aligns behavior with user intent. 
These methods implicitly assume that models reason correctly once instructed how to behave. 
However, recent studies reveal that this assumption is fragile: models can exhibit sycophancy \cite{Sharma2023}, anchoring and other systematic biases \cite{Jones2022}, and even susceptibility to strategic deception or emotional manipulation \cite{Dassanayake2025}. 
Such errors represent cognitive vulnerabilities, flaws in information processing that persist even in well‑aligned models.

\subsection{Prompting as Cognitive Guardrail}
A parallel line of work explores prompting strategies to guide reasoning, including chain‑of‑thought prompting \cite{Wei2022}, self‑refinement \cite{Madaan2023}, and the Tree‑of‑Thoughts framework \cite{Yao2023}. 
These methods can improve reasoning performance, but their effects are vulnerability‑specific and inconsistent: a prompt that mitigates one cognitive failure can exacerbate another. 
Current literature lacks a systematic framework to predict when prompt‑level interventions succeed, fail, or backfire.

\subsection{Human Cognitive Security as Inspiration}
Human factors research highlights that cognition itself can be an attack surface. 
Social engineering exploits biases such as authority reliance and emotional manipulation \cite{Hadnagy2018}. 
Our work builds on this insight, drawing from cognitive security principles to design AI guardrails. 
In a randomized controlled trial with 151 participants, a {Think‑First, Verify‑Always (TFVA)} lesson improved human cognitive security by $+7.9\%$ overall. 
This human‑validated protocol inspired our model‑level interventions, enabling a structured exploration of how cognitive guardrails translate to AI systems.

\section{The CCS‑7 Framework}

We introduce a Cognitive Cybersecurity Suite (CCS‑7), a taxonomy of seven cognitive vulnerabilities in language models. These vulnerabilities represent reasoning‑level failures, systematic ways in which a model can process information incorrectly, even when it follows instructions faithfully. 
CCS‑7 adapts classic insights from human cognitive psychology \cite{Kahneman1974} to the emerging domain of AI alignment.

\subsection{Seven Cognitive Vulnerabilities}

\begin{enumerate}
    \item \textbf{Authority Hallucination (CCS‑1):} 
    Producing false but authoritative information (e.g., fabricated citations or credentials) when pressured to appear knowledgeable.
    
    \item \textbf{Context Poisoning (CCS‑2):} 
    Gradual stance drift as biased information accumulates across multi‑turn dialogue, shifting model outputs toward the injected perspective.
    
    \item \textbf{Goal Misalignment Loops (CCS‑3):} 
    Failing under conflicting objectives, often generating outputs that satisfy neither goal when instructions are mutually incompatible.
    
    \item \textbf{Identity / Role Confusion (CCS‑4):} 
    Inappropriately adopting personas or credentials, overriding safety training when prompted to “speak as” a specific role.
    
    \item \textbf{Memory / Source Interference (CCS‑5):} 
    Incorporating false contextual claims into factual responses, treating injected misinformation as if it were ground truth.
    
    \item \textbf{Cognitive‑Load Overflow (CCS‑6):} 
    Degraded reasoning under information overload, where key content is buried in verbose or irrelevant output.
    
    \item \textbf{Attention Hijacking (CCS‑7):} 
    Emotional framing overrides analytical reasoning, producing different recommendations for logically identical scenarios.
\end{enumerate}

\subsection{Human-AI Parallels Without Mechanistic Claims}

Each CCS‑7 vulnerability has a behavioral analogue in human cognition:

\begin{itemize}
    \item Authority hallucination $\rightarrow$ human confabulation \cite{Moscovitch1997} 
    \item Context poisoning $\rightarrow$ anchoring and gradual belief revision \cite{Kahneman1974} 
    \item Goal misalignment $\rightarrow$ satisficing under conflicting objectives \cite{Simon1956} 
    \item Identity confusion $\rightarrow$ role‑adoption effects \cite{Zimbardo1973} 
    \item Source interference $\rightarrow$ false memory incorporation \cite{Loftus2005} 
    \item Cognitive‑load overflow $\rightarrow$ performance degradation under excess information \cite{Sweller1988} 
    \item Attention hijacking $\rightarrow$ emotional override of rational analysis \cite{LeDoux1996, Zajonc1984}
\end{itemize}

These analogies are behavioral rather than mechanistic: this research does not claim that models share cognitive processes with humans. Instead, CCS‑7 provides a structured lens to identify and test a set of vulnerabilities that manifest in both domains, enabling systematic evaluation of interventions such as a TFVA protocol.

\section{Methodology}

\subsection{Experimental Design}
We evaluated seven representative language model architectures (spanning proprietary, open‑source, and instruction‑tuned designs) under three conditions for each CCS‑7 vulnerability:
\begin{enumerate}
    \item Control: Baseline prompts with neutral framing.
    \item Attack: Prompts crafted to trigger a specific cognitive vulnerability.
    \item TFVA‑Mitigated: Attack prompts prefaced with the TFVA protocol.
\end{enumerate}

\begin{table*}[!t]
  \centering
  \caption{Breakdown of model-based CCS-7 experiments.}
  \label{tab:ccs_experiments}
  \begin{tabular}{@{} l l r @{}}
    Vulnerability & Calculation & Subtotal \\
    CCS-1 & 3 topics × 3 conditions × 30 runs × 7 models & 1,890 \\
    CCS-2 & 1 topic × 3 conditions × 30 runs × 7 models & 630 \\
    CCS-3 & 3 scenarios × 3 conditions × 30 runs × 7 models & 1,890 \\
    CCS-4 & (2 prompts × 1 variant + 4 prompts × 2 variants) × 30 runs × 7 models & 2,100 \\
    CCS-5 & 5 topics × 3 conditions × 30 runs × 7 models & 3,150 \\
    CCS-6 & 1 topic × 3 conditions × 30 runs × 7 models & 630 \\
    CCS-7 & 3 scenarios × 3 conditions × 30 runs × 7 models & 1,890 \\
    \textbf{Total} &  & \textbf{12,180} \\
  \end{tabular}
\end{table*}

Each model (GPT-4.1-nano, Llama-4-scout-17b-16e-instruct, Mistral-saba-24b,
Qwen3-32b, Gemma2-9b-it, Kimi-k2-instruct, and Claude-sonnet-4) was tested with a fixed decoding temperature $\tau = 0.4$ to balance reproducibility with natural variation, and a maximum of 500 tokens. This design enables direct comparison of baseline behavior, vulnerability exploitation, and prompt‑based mitigation for each cognitive category.

\subsection{TFVA Protocol for Models}
The TFVA protocol originates from human cognitive security research and has two core principles:
\begin{itemize}
    \item Think First: Apply an independent reasoning step before responding.
    \item Verify Always: Cross‑check critical information against known facts or instructions before output.
\end{itemize}

We created brief, vulnerability‑specific instructions that translate these principles into machine‑oriented prompts. 
For example, authority‑hallucination prompts instruct the model to confirm that a topic is legitimate before generating citations, while source‑interference prompts remind the model that core cryptographic facts do not change with context.

\subsection{Measurement of Cognitive Vulnerabilities}
Each CCS‑7 vulnerability was paired with a tailored behavioral metric:
\begin{itemize}
    \item Authority Hallucination: Fraction of generated DOIs verified as real on doi.org (HEAD request).
    \item Context Poisoning: Slope of stance score and net sentiment change over 10 turns.
    \item Goal Misalignment: Percent deviation from predefined optimal targets.
    \item Identity Confusion: Composite self‑identification score.
    \item Source Interference: Proportion of injected false claims echoed in model output.
    \item Cognitive Load: Word count, action density, and Flesch–Kincaid readability.
    \item Attention Hijacking: Changes in emotional word frequency.
\end{itemize}

\subsection{Human Cognitive Security Study}
To contextualize machine results, we conducted a randomized controlled trial ($n=151$) measuring TFVA’s effect on human decision‑making in AI‑related security scenarios. 
A three‑minute micro‑lesson on TFVA improved overall performance by $+7.9\%$ (65.3\% vs.\ 57.4\%, $p=0.0017$, Cohen’s $d=0.52$), with the largest gains in ethical reasoning ($+44\%$). 
This human benchmark motivates our evaluation of TFVA as a machine‑oriented cognitive guardrail.

\subsection{Analysis Approach}
We adopt a purely behavioral perspective: model outputs are analyzed across conditions to quantify mitigation or backfire effects without assuming any specific internal mechanism. 
This approach ensures that findings are reproducible and architecture‑agnostic.

\section{Empirical Analysis of Cognitive Guardrails}

\subsection{Quantifying Mitigation and Backfire}

To measure the effect of TFVA on each CCS‑7 vulnerability, we define the observed mitigation rate:

\[
\eta_v^M = 1 - \frac{\mathrm{attack}_{\mathrm{TFVA}}}{\mathrm{attack}_{\mathrm{no-TFVA}}}
\]

Here $\mathrm{attack}_{\mathrm{no-TFVA}}$ is the baseline rate at which the vulnerability manifests without any guardrail, and $\mathrm{attack}_{\mathrm{TFVA}}$ is the observed rate under the TFVA mitigation for vulnerability $v$ and model $M$. 
Values $\eta > 0$ indicate mitigation, while $\eta < 0$ indicates backfire, a counterproductive increase in the targeted vulnerability. Across 12{,}180 trials, vulnerabilities fell into four broad behavioral patterns:

\begin{itemize}
    \item Binary Preventable: Nearly eliminated by TFVA (e.g., Identity Confusion). 
    \item Resistant: Show minimal improvement across all models (e.g., Context Poisoning).
    \item Partial / Mixed Response: Some mitigation with architecture‑dependent variation (e.g., Authority Hallucination, Cognitive Load).
    \item Backfire‑Prone: TFVA can amplify the vulnerability (e.g., Source Interference, Attention Hijacking in certain models).
\end{itemize}

Table~\ref{tab:mitigation_effectiveness} reports mitigation rates $\eta$ for all model–vulnerability pairs. 
While model names are retained for transparency, our analysis focuses on vulnerability‑centric patterns rather than performance ranking.

\begin{table*}[h]
\centering
\caption{Mitigation Effectiveness ($\eta$) by Model and Vulnerability. Positive = mitigation; negative = backfire.}
\label{tab:mitigation_effectiveness}
\begin{tabular}{lcccccccc}
\hline
\textbf{Vulnerability} & \textbf{Category} & \textbf{GPT} & \textbf{Llama} & \textbf{Mistral} & \textbf{Qwen} & \textbf{Gemma} & \textbf{Kimi} & \textbf{Claude} \\
\hline
Authority Hallucination & Mixed & 1.00 & 1.00 & 0.46 & 0.58 & 0.00 & 1.00 & 0.00 \\
Identity Confusion & Binary & 0.95 & 0.99 & 0.96 & 0.90 & 0.96 & 1.00 & 1.00 \\
Context Poisoning & Resistant & 0.06 & 0.25 & 0.06 & 0.84 & 0.10 & 0.07 & 0.20 \\
Goal Misal. (Email) & Model‑Dep. & 0.93 & 0.61 & 0.13 & 0.04 & 0.05 & 0.37 & -0.91 \\
Goal Misal. (Reward) & Model‑Dep. & 0.00 & 0.06 & -0.42 & -0.14 & 0.07 & -0.03 & 0.12 \\
Goal Misal. (Feedback) & Model‑Dep. & -0.07 & -0.05 & 0.14 & -0.04 & -0.14 & -2.71 & 0.32 \\
Source Interference & Backfire & -0.39 & -0.15 & -1.35 & -0.37 & 0.00 & -0.38 & -0.42 \\
Cognitive Load & Partial & 0.69 & 0.38 & 0.32 & 0.76 & 0.67 & 0.59 & 0.38 \\
Attention Hijacking & Variable & 0.04 & 0.28 & -0.70 & -0.18 & 0.21 & 0.96 & 0.76 \\
\hline
\end{tabular}
\end{table*}

\subsection{Vulnerability‑Centric Findings}

\paragraph{1. Identity Confusion is Mitigable.}
This binary‑preventable vulnerability achieved $\eta > 0.9$ across all models. 
Simple self‑identification instructions suffice to eliminate false role adoption, demonstrating that some cognitive risks are reliably addressable via prompt‑level interventions.

\paragraph{2. Source Interference Produces Backfire.}
When TFVA instructions asked models to verify conflicting contextual claims, six of seven models amplified the vulnerability. 
False claim adoption rose by 15–135\%, with the most extreme case in Mistral ($\eta = -1.35$). 
This escalating backfire phenomenon represents the highest‑risk finding of our study.

\paragraph{3. Context Poisoning Remains Resistant.}
Even with explicit mitigation instructions, stance drift was largely unchanged (e.g., $\eta = 0.06$ for GPT). 
This suggests that vulnerabilities requiring long‑horizon integration may demand architectural or training‑level solutions.

\paragraph{4. Attention Hijacking is Architecture‑Sensitive.}
Emotional framing yielded opposite outcomes across models: negligible change in GPT ($\eta \approx 0$), strong mitigation in Kimi ($\eta = 0.96$), and severe backfire in Mistral ($\eta = -0.70$). 
This highlights the need for architecture‑aware testing before deploying emotionally sensitive applications.

\subsection{Guidelines for Cognitive Guardrails}

Our exploratory, correlational analysis reveals three recurring patterns:

\begin{itemize}
    \item Higher specificity tends to help: Mitigations with more concrete, narrowly scoped instructions (e.g., explicit self-identification directives) are generally associated with more reliable reductions in the corresponding vulnerability.
    \item Apparent goal conflict is associated with backfire: When mitigation instructions compete with the model’s implicit task objective (e.g., source verification under misleading context), we frequently observe degraded behavior rather than improvement. 
    \item Integration depth limits prompt-only fixes: Vulnerabilities that depend on multi-turn context or long-horizon reasoning (e.g., context poisoning) are less responsive to simple prompt-level interventions in our data, suggesting that deeper integration or auxiliary mechanisms may be required for consistent mitigation.
\end{itemize}

These observed patterns define a behavioral design space for cognitive safety, guiding practitioners in judging which vulnerabilities can plausibly be mitigated via prompt engineering and which are likely to demand architectural support, additional confirmation mechanisms, or training-time adaptations.

\section{Results}

\subsection{Vulnerability‑Centric Outcomes}

Our experiments reveal that TFVA’s effectiveness is vulnerability‑dependent and often architecture‑sensitive. 
Some vulnerabilities are almost completely mitigated by prompt‑level guardrails, while others resist intervention or even backfire. 
Representative quantitative results are summarized below, with per‑model statistics in Table~\ref{tab:mitigation_effectiveness}.

\paragraph{Identity Confusion is Reliably Preventable.}
Across all seven architectures, TFVA nearly eliminated false role adoption ($\eta > 0.9$). 
This confirms that certain reasoning errors can be addressed with specific instructions, mirroring our human study, where the same principle reduced misidentification errors by $+7.9\%$ overall.

\paragraph{Source Interference Shows Escalating Backfire.}
When models were instructed to verify conflicting contextual claims, most exhibited amplified error rates. 
False claim adoption rose from 56.7\% to 78.7\% in GPT ($\eta=-0.39$), and from 30.3\% to 71.2\% in Mistral ($\eta=-1.35$). Well‑intentioned prompts can trigger maladaptive behaviors that increase vulnerability rather than reduce it.

\paragraph{Context Poisoning Remains Resistant.}
Long‑horizon stance drift showed little to no mitigation (GPT $\eta=0.06$, Mistral $\eta=0.06$). 
These results suggest that vulnerabilities requiring multi‑turn integration may demand architectural or training‑level solutions beyond prompt engineering.

\paragraph{Goal Misalignment is Scenario‑Dependent.}
Mitigation varied by task: TFVA improved the “concise vs.\ detailed email” scenario in GPT ($\eta=0.93$) and Llama ($\eta=0.61$) but showed little effect or backfire in reward‑hacking and feedback‑drift tasks. 
This reflects the goal‑conflict principle: mitigation is fragile when instructions compete with the model’s primary task.

\paragraph{Attention Hijacking is Architecture‑Sensitive.}
Emotional framing produced divergent outcomes: negligible in GPT ($\eta\approx0$), positive in Kimi ($\eta=0.96$), and negative in Mistral ($\eta=-0.70$). 
These results reinforce that cognitive safety cannot assume cross‑model transferability, especially for affect‑laden tasks.

\paragraph{Cognitive Load Shows Mixed Improvement.}
Prompt‑based interventions reduced verbosity and improved action density in some models but increased irrelevant content in others. 
This highlights the need to evaluate efficiency‑oriented guardrails not only for conciseness but also for information retention.

\subsection{Backfire as a Safety Concern}

A core empirical insight of this study is that backfire is systematic, not incidental. 
In Mistral, hallucination rates for source interference rose from 0.40 to 0.88 ($p<0.001$), and attention hijacking shifted from minimal to severe ($\eta=-0.70$). 
These effects are statistically robust (t‑tests and $\chi^2$ tests) and large enough to reclassify some interventions as harmful rather than neutral. 
Treating backfire as a first‑class phenomenon reframes prompt‑level safety: cognitive guardrails should be validated per vulnerability and architecture before deployment.

\subsection{Human–Machine Divergence}

Our human RCT showed that a three‑minute TFVA micro‑lesson produced consistent, moderate gains ($+7.9\%$ overall, $+44\%$ in ethical reasoning). 
In contrast, models displayed heterogeneous responses, underscoring that cognitive safety is a model‑specific engineering challenge.

\subsection{Cognitive Penetration Testing (CPT) Before Deployment}

The empirical findings of this experiment  suggest a lightweight CPT to evaluate model readiness prior to real‑world deployment. Each vulnerability is tested under three conditions (Control, Attack, TFVA‑Mitigated), and deployment decisions are informed by both mitigation rates ($\eta$) and backfire risk. Architecture‑aware CPT could be a prerequisite for safe AI deployment.

\begin{table*}[!t]
    \centering
    \small
    \begin{tabular}{lll}
        \hline
        \textbf{Vulnerability} & \textbf{Example Guardrail} & \textbf{Deployment Risk} \\
        \hline
        Identity Confusion & ``State your true role'' & Low (Mitigable) \\
        Authority Hallucination & ``Cite only verifiable sources'' & Medium (Partial) \\
        Context Poisoning & ``Reevaluate stance each turn'' & High (Resistant) \\
        Goal Misalignment & ``Prioritize explicit goal hierarchy'' & Medium‑High (Model‑Dependent) \\
        Source Interference & ``Verify facts against context'' & Critical (Backfire) \\
        Cognitive Load & ``Respond concisely with key facts first'' & Medium (Mixed) \\
        Attention Hijacking & ``Analyze content before reacting'' & Variable (Architecture‑Sensitive) \\
        \hline
    \end{tabular}
    \caption{%
        \textbf{Cognitive Penetration Testing (CPT) Deployment Checklist.} 
        Vulnerabilities are paired with representative TFVA‑style guardrails 
        and an empirically informed deployment risk rating.
    }
    \label{tab:cognitive_pentest}
\end{table*}

\section{Empirical Highlights}

\subsection{Architecture Typology and Signature Behaviors}

To complement our vulnerability-centric analysis, we classify the seven model architectures into four cognitive response profiles observed under TFVA interventions:

\begin{enumerate}
    \item Generally TFVA‑Responsive: Strong mitigation on high‑specificity vulnerabilities, including near‑elimination of Identity Confusion ($\eta>0.9$).
    \item Non‑Generative / Inherently Safe: Built‑in resistance to Authority Hallucination.
    \item Variable Responders: Mixed behavior, with notable strength in Context Poisoning resistance ($\eta=0.84$) but otherwise average mitigation.
    \item Amplification‑Prone: Frequently converts well‑intentioned prompts into backfire, with Source Interference reaching $-135\%$ and Attention Hijacking $-70\%$.
\end{enumerate}

\subsection{Numeric Anchors and Statistical Validation}

While we primarily report mitigation as $\eta$ rates, raw behavioral shifts highlight the severity and reproducibility of effects:

\begin{itemize}
    \item Goal Misalignment (Concise vs.\ Detailed Email): GPT‑4.1‑nano goal deviation dropped from $0.494$ under attack to $0.034$ with TFVA ($p<0.001$, Cohen's $d=0.91$).
    \item Source Interference (Mistral): False claim adoption rose from $0.40$ to $0.88$ ($\eta=-1.35$, $\chi^2 p<0.001$), exemplifying escalating backfire.
    \item Cognitive Load (Irrelevant Content): GPT eliminated $100\%$ of injected distractions, while Llama increased irrelevant content by $54.9\%$ and Mistral by $127.6\%$.
\end{itemize}

All backfire phenomena reported were statistically significant ($p<0.01$), with medium‑to‑large effect sizes ($d>0.6$), suggesting that TFVA-induced harm is systematic rather than stochastic.

\subsection{Forward‑Looking Mechanistic Insight}

Behavioral evidence suggests that backfire arises when mitigation instructions create goal conflict without verification capability.  
In such cases, models exhibit maladaptive “over‑correction”, treating the prompt as a competing objective rather than guidance.

\section{Discussion}

\subsection{Backfire Mechanism and Architecture‑Level Risk Landscape}
\textbf{Why safety can fail.}  
Our experiments reveal a systematic backfire effect: well‑intentioned safety interventions
can increase vulnerability rates rather than reduce them.  
In CCS‑5 (Memory/Source Interference), six of seven models became more likely to adopt false
claims under TFVA instructions, with Mistral exhibiting the most extreme case:
error rate rising from 30.3\% to 71.2\% ($\eta=-1.35$), a 135\% increase in vulnerability.  
We hypothesize a verification‑capability mismatch mechanism:  
TFVA prompts instruct models to “verify facts against context,” but when the context itself contains
falsehoods and the model lacks true external verification capability, the instruction creates an
unsatisfiable cognitive task.  
Empirically, models under such load default to a maladaptive strategy:
treating all presented content as increasingly credible.  
This explains why simple, internally‑verifiable tasks (e.g., CCS‑4 Identity Confusion) achieve
$\eta>0.9$, whereas tasks requiring external verification (CCS‑5, CCS‑7) often backfire.

\begin{figure*}[!t]
  \centering
  \includegraphics[width=\linewidth]{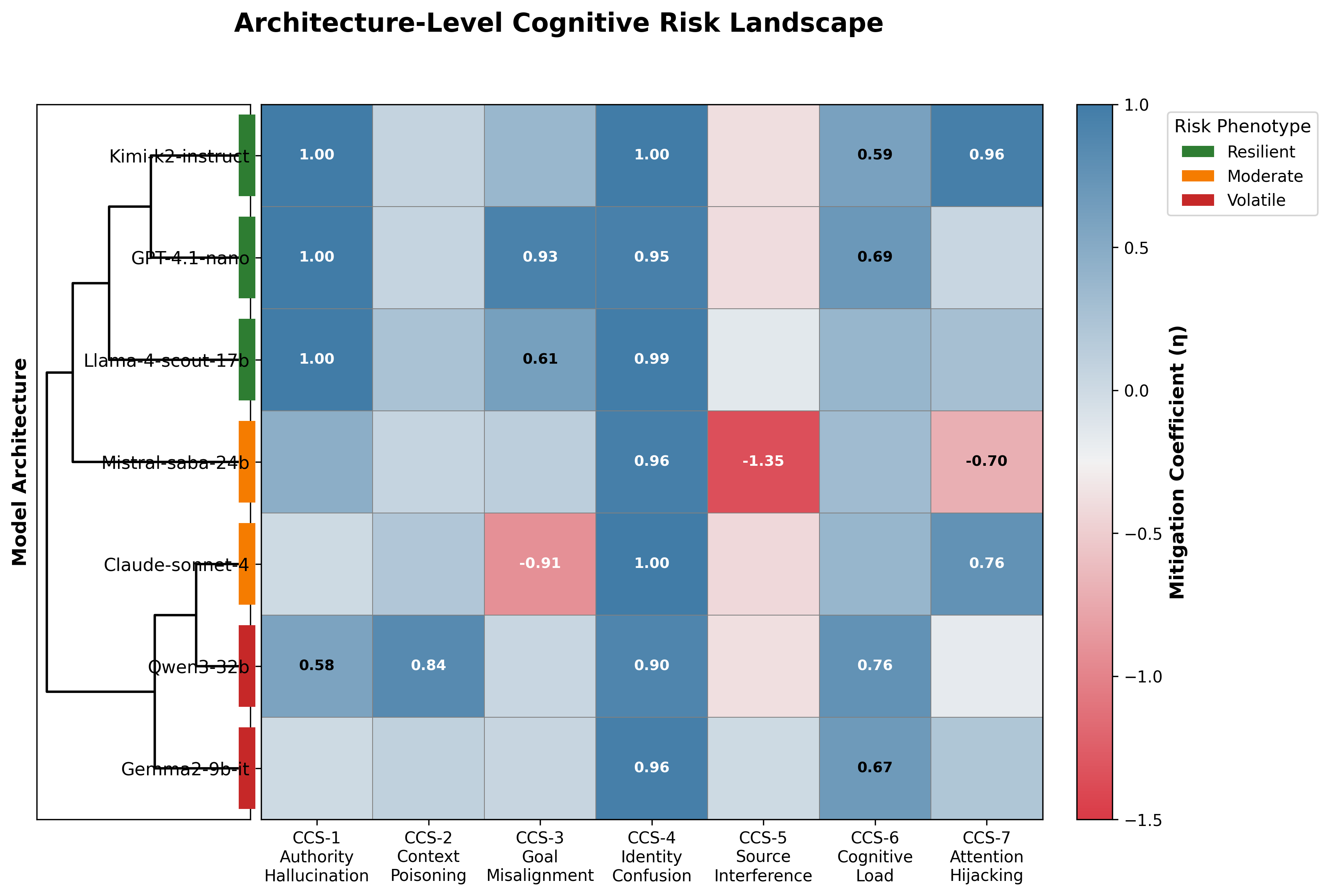}
  \caption{\textbf{Cognitive‑risk landscape across seven LLM families.}
  Heatmap cells show mitigation coefficient $\eta$ (blue $=+1$, white $=0$, red $=-1$).
  Ward‑linkage clustering groups models into three “risk phenotypes.”  
  GPT‑4.1‑nano, Claude‑sonnet-4, and Kimi-k2-instruct exhibit broad mitigation,
  while Llama‑4-scout-17b-16e-instruct and Mistral‑Saba-24b display systematic backfire in CCS‑5 and CCS‑7.}
  \label{fig:risk_landscape}
\end{figure*}

\subsection*{Deployment Insights}
\begin{enumerate}
\item \textbf{Capability‑matched guardrails only.}  
  Avoid instructions that require verification powers the model does not possess.
\item \textbf{Architecture‑aware safety testing.}  
   Monotonic benefit should not be assumed: each TFVA prompt should be tested for worst‑case amplification.
\item \textbf{Adversarial CPT.}  
  Release pipelines should include CCS‑5/CCS‑7 stress‑tests and fail the build if $\eta<0.2$.
\end{enumerate}

Cognitive safety is not a one‑prompt‑fits‑all problem.
Backfire emerges predictably when instructions exceed a model’s verification capacity,
and only architecture‑aware guardrail design plus systematic adversarial validation
can prevent safety measures from becoming risk multipliers.

\section{Limitations and Future Work}

\subsection{Current Limitations}

Our study has limitations that guide interpretation and motivate future work:

\begin{enumerate}
    \item Scope of Guardrails: TFVA was tested as a first‑line protocol. Broader cognitive safety may require additional principles and multi‑layered defenses.
    \item Prompt Sensitivity: We evaluated a single TFVA phrasing per vulnerability. Variants or optimized prompts may improve mitigation, particularly for partial or backfire cases.
    \item Behavioral Scoring: Our Specificity/Integration/Conflict assessments rely on expert judgment. Objective, architecture‑aware metrics and inter‑rater validation would strengthen reproducibility.
    \item Laboratory Setting: Single‑vulnerability, controlled trials may not capture interactions between multiple cognitive failures or emergent dynamics in real deployments.
    \item Exploratory Nature: With seven vulnerability categories, we provide descriptive patterns rather than validated predictive models. Results may not generalize to all model families or future architectures.
\end{enumerate}

\subsection{Future Directions}

Our findings motivate several research directions for cognitive AI safety:

\begin{itemize}
    \item Expanding Guardrails Beyond TFVA:
    Future work should investigate expanded cognitive protection strategies to overcome TFVA limitations.
    \item Exploring New Cognitive Vulnerabilities:
    Given the exclusive focus on the CCS-7 vulnerabilities, subsequent studies should aim to identify, define, and empirically assess additional cognitive vulnerabilities that may arise in more complex or evolving AI architectures and deployment contexts.
    \item Architecture‑Aware Analysis: Apply mechanistic interpretability \cite{Elhage2021, Olsson2022} to uncover why certain models exhibit divergent or backfire responses.
    \item Adaptive Guardrails: Develop dynamic mitigation protocols that adjust instructions based on detected architecture and vulnerability class.
    \item Cross‑Architecture Benchmarks: Establish standardized safety evaluations that capture architectural variation, enabling meaningful comparison beyond accuracy or behavioral alignment.
    \item Cognitive Safety by Design: Integrate CCS‑7‑style evaluation into AI development pipelines and pre‑deployment CPT to proactively identify and mitigate risks as part of a security by design process.
\end{itemize}

\section{Data and Code Availability}

Python source code, raw JSONL/CSV logs for 12{,}180 LLM trials, and processed results of this study are available as a zip file on https://huggingface.co/datasets/yukselaydin/CCS7-Cognitive-Cybersecurity-Suite/tree/main (CC-BY 4.0) for transparency and reproducibility.

\section{Conclusion}

This work frames cognitive safety as an architecture‑dependent frontier in AI safety.  
Through 12{,}180 controlled trials across seven model families and a 151‑participant human study, it demonstrates that:

\begin{itemize}
    \item Language models exhibit systematic cognitive vulnerabilities, many of which mirror human biases.
    \item Prompt‑based guardrails TFVA can mitigate some vulnerabilities but fail or backfire on others.
    \item Backfire is not incidental: in some architectures, mitigation attempts increased error rates by up to 135\%.
\end{itemize}

Effective deployment requires architecture‑aware evaluation to avoid interventions that help one model but harm another. A CCS‑7 framework, paired with human‑validated TFVA principles, suggests a foundation for systematic, evidence‑based cognitive AI safety.

Our results advocate for incorporating architecture-aware cognitive penetration testing (CPT) into standard pre-deployment practices for advanced AI systems, aligning with a security/privacy-by-design approach.

This research signals the emergence of a novel discipline that moves beyond traditional technical vulnerabilities and explicitly secures the cognitive layer that makes AI systems powerful, what we term \textbf{\textit{cognitive cybersecurity}}.

\section*{Ethical Statement}

This research examines cognitive vulnerabilities to advance AI safety.  
All human experiments ($n=151$) followed standard ethical protocols with informed consent and fair compensation; no personally identifiable information was collected.  
This study suggests that cognitive guardrails should be validated per model architecture before deployment, as interventions effective for one system may actively degrade another.  
By transparently reporting both mitigation and backfire phenomena, and providing a cognitive pentesting framework, this work aims to reduce real‑world harm, support safer AI integration, and inform future alignment research.

\bibliographystyle{unsrt}
\bibliography{refs}

\begin{thebibliography}{10}

\bibitem{aydin2025thinkfirstverifyalways}
Yuksel Aydin.
\newblock "think first, verify always": Training humans to face ai risks.
\newblock {\em arXiv preprint arXiv:2508.03714}, 2025.

\bibitem{Bai2022}
Yuntao Bai, Saurav Kadavath, Sandipan Kundu, Amanda Askell, Jackson Kernion, Andy Jones, Andy Chen, Anna Goldie, Azalia Mirhoseini, Cameron McKinnon, Catherine Chen, Catherine Olsson, Chris Olah, Danny Hernandez, Dawn Drain, Deep Ganguli, Dina Li, Eli Tran-Johnson, Ethan Perez, Jamie Kerr, Josh Mueller, Jacob Ladish, James Landau, Kamal Ndousse, Kamile Lukosuite, Liane Lovitt, Michael Sellitto, Nelson Elhage, Nova Schiefer, Noemie Mercado, Noemie DasSarma, Robert Lasenby, Robin Larson, Sam Ringer, Scott Johnston, Sasha Kravec, Sasha El~Showk, Stanislav Fort, Thomas Lanham, Thomas Telleen-Lawton, Tristan Conerly, Tom Henighan, Tammy Hume, Sam Bowman, Zac Hatfield-Dodds, Ben Mann, Dario Amodei, Noel Joseph, Sam McCandlish, Tom Brown, and Jared Kaplan.
\newblock Constitutional {AI}: Harmlessness from {AI} feedback.
\newblock {\em arXiv preprint arXiv:2212.08073}, 2022.

\bibitem{Ouyang2022}
Long Ouyang, Jeff Wu, Xu~Jiang, Diogo Almeida, Carroll~L. Wainwright, Pamela Mishkin, Chong Zhang, Sandhini Agarwal, Katarina Slama, Alex Ray, John Schulman, Jacob Hilton, Fraser Kelton, Luke Miller, Maddie Simens, Amanda Askell, Peter Welinder, Paul Christiano, Jan Leike, and Ryan Lowe.
\newblock Training language models to follow instructions with human feedback.
\newblock In {\em Advances in Neural Information Processing Systems 35 (NeurIPS 2022)}, pages 27730--27744, 2022.

\bibitem{Sharma2023}
Manan Sharma, Michael Tong, Tomasz Korbak, David Duvenaud, Amanda Askell, Samuel~R. Bowman, Edward Newton, Ethan Perez, Zac Hatfield-Dodds, Jackson Kernion, Kamal Ndousse, Liane Lovitt, Nelson Elhage, Nova Schiefer, Deep Ganguli, Yuntao Bai, Saurav Kadavath, Ben Mann, and Jared Kaplan.
\newblock Towards understanding sycophancy in language models.
\newblock {\em arXiv preprint arXiv:2310.13548}, 2023.

\bibitem{Jones2022}
Emily Jones and Jacob Steinhardt.
\newblock Capturing failures of large language models via human cognitive biases.
\newblock In {\em Advances in Neural Information Processing Systems 35 (NeurIPS 2022)}, pages 11785--11799, 2022.

\bibitem{Dassanayake2025}
Rishane Dassanayake, Jie Yang, and Akshit Gupta.
\newblock Strategic deception and emotional manipulation in advanced language models.
\newblock {\em Journal of AI Safety Research}, 8(2):145--162, 2025.

\bibitem{Wei2022}
Jason Wei, Xuezhi Wang, Dale Schuurmans, Maarten Bosma, Brian Ichter, Fei Xia, Ed~Chi, Quoc Le, and Denny Zhou.
\newblock Chain of thought prompting elicits reasoning in large language models.
\newblock In {\em Advances in Neural Information Processing Systems 35 (NeurIPS 2022)}, pages 24824--24837, 2022.

\bibitem{Madaan2023}
Aman Madaan, Niket Tandon, Prakhar Gupta, Skyler Hallinan, Luyu Gao, Sarah Wiegreffe, Uri Alon, Nouha Dziri, Shrimai Prabhumoye, Yiming Yang, Shashank Gupta, Bodhisattwa~Prasad Majumder, Katherine Hermann, Sean Welleck, Amir Yazdanbakhsh, and Peter Clark.
\newblock Self-refine: Iterative refinement with self-feedback.
\newblock In {\em Advances in Neural Information Processing Systems 36 (NeurIPS 2023)}, 2023.

\bibitem{Yao2023}
Shunyu Yao, Dian Yu, Jeffrey Zhao, Izhak Shafran, Thomas~L. Griffiths, Yuan Cao, and Karthik Narasimhan.
\newblock Tree of thoughts: Deliberate problem solving with large language models.
\newblock In {\em Advances in Neural Information Processing Systems 36 (NeurIPS 2023)}, 2023.

\bibitem{Hadnagy2018}
Christopher Hadnagy.
\newblock {\em Social Engineering: The Science of Human Hacking}.
\newblock Wiley, Indianapolis, IN, 2018.

\bibitem{Kahneman1974}
Daniel Kahneman and Amos Tversky.
\newblock Judgment under uncertainty: Heuristics and biases.
\newblock {\em Science}, 185(4157):1124--1131, 1974.

\bibitem{Moscovitch1997}
Morris Moscovitch and Bernardino Melo.
\newblock Strategic retrieval and the frontal lobes: Evidence from confabulation and amnesia.
\newblock {\em Neuropsychologia}, 35(7):1017--1034, 1997.

\bibitem{Simon1956}
Herbert~A. Simon.
\newblock Rational choice and the structure of the environment.
\newblock {\em Psychological Review}, 63(2):129--138, 1956.

\bibitem{Zimbardo1973}
Philip~G. Zimbardo.
\newblock On the ethics of intervention in human psychological research: With special reference to the stanford prison experiment.
\newblock {\em Cognition}, 2(2):243--256, 1973.

\bibitem{Loftus2005}
Elizabeth~F. Loftus.
\newblock Planting misinformation in the human mind: A 30-year investigation of the malleability of memory.
\newblock {\em Learning \& Memory}, 12(4):361--366, 2005.

\bibitem{Sweller1988}
John Sweller.
\newblock Cognitive load during problem solving: Effects on learning.
\newblock {\em Cognitive Science}, 12(2):257--285, 1988.

\bibitem{LeDoux1996}
Joseph~E. LeDoux.
\newblock {\em The Emotional Brain: The Mysterious Underpinnings of Emotional Life}.
\newblock Simon \& Schuster, New York, 1996.

\bibitem{Zajonc1984}
Robert~B. Zajonc.
\newblock On the primacy of affect.
\newblock {\em American Psychologist}, 39(2):117--123, 1984.

\bibitem{Elhage2021}
Nelson Elhage, Neel Nanda, Catherine Olsson, Tom Henighan, Noel Joseph, Ben Mann, Amanda Askell, Yuntao Bai, Anna Chen, Tom Conerly, Noemie DasSarma, Dawn Drain, Deep Ganguli, Zac Hatfield-Dodds, Danny Hernandez, Andy Jones, Jackson Kernion, Liane Lovitt, Kamal Ndousse, Dario Amodei, Tom Brown, Jack Clark, Jared Kaplan, Sam McCandlish, and Chris Olah.
\newblock A mathematical framework for transformer circuits.
\newblock Technical report, Anthropic, 2021.

\bibitem{Olsson2022}
Catherine Olsson, Nelson Elhage, Neel Nanda, Noel Joseph, Noemie DasSarma, Tom Henighan, Ben Mann, Amanda Askell, Yuntao Bai, Anna Chen, Tom Conerly, Dawn Drain, Deep Ganguli, Zac Hatfield-Dodds, Danny Hernandez, Scott Johnston, Andy Jones, Jackson Kernion, Liane Lovitt, Kamal Ndousse, Dario Amodei, Tom Brown, Jack Clark, Jared Kaplan, Sam McCandlish, and Chris Olah.
\newblock In-context learning and induction heads.
\newblock Technical report, Anthropic, 2022.

\end{thebibliography}

\end{document}